\documentclass{aa}
\usepackage{epsfig}

\newcommand{\msun}{\mbox{$M_{\odot}$}}
\newcommand{\Msun}{\mbox{$M_{\odot}$}}
\newcommand{\lsun}{\mbox{$L_{\odot}$}}

\newcommand{\teff}{\mbox{$T_{\rm eff}$}}

\newcommand{\vinf}{\mbox{$v_{\infty}$}}

\newcommand{\ratio}{\mbox{$v_{\infty}$/$v_{\rm esc}$}}
\newcommand{\mdot}{\mbox{$\dot{M}$}}

\newcommand{\msunyr}{\mbox{$M_{\odot} {\rm yr}^{-1}$}}

\newcommand{\logll}{\mbox{$\log (L/L_{\odot})$}}

\begin{document}


\title{Predictions of variable mass loss for Luminous Blue Variables}

\author{Jorick S. Vink\inst{1,2}
\and Alex de Koter\inst{3}}

\offprints{Jorick S. Vink, j.vink@ic.ac.uk}

\institute{Imperial College, Blackett Laboratory, 
           Prince Consort Road, London, SW7 2BZ, U.K.
           \and
            Astronomical Institute, Utrecht University,
            P.O.Box 80000, NL-3508 TA Utrecht, The Netherlands
            \and
            Astronomical Institute ``Anton Pannekoek'', University of Amsterdam,
            Kruislaan 403, NL-1098 SJ Amsterdam, The Netherlands
}

\date{Received 16 April 2002; accepted 8 July 2002}

\titlerunning{Predictions of variable mass loss for LBVs}
\authorrunning{Jorick S. Vink \& Alex de Koter}

\abstract{We present radiation-driven wind models for Luminous Blue 
Variables (LBVs) and predict their mass-loss rates. We study the effects of 
lower masses and modified abundances in comparison to the normal OB supergiants, 
and we find that the main difference in mass loss is due to the lower masses of LBVs. 
In addition, we find that the increase in helium abundance changes the 
mass-loss properties by small amounts (up to about 0.2 dex in log $\dot{M}$), 
while CNO processing is relatively unimportant for the mass-loss rate.
A comparison between our mass loss predictions 
and the observations is performed for four relatively 
well-studied LBVs. The comparison 
shows that (i) the winds of LBVs are driven by radiation pressure 
on spectral lines, (ii) the variable mass loss behaviour of 
LBVs during their S\,Doradus-type variation cycles is explained 
by changes in the line driving efficiency, notably due to the 
recombination/ionisation of Fe\,{\sc iv}/{\sc iii} and 
Fe\,{\sc iii}/{\sc ii}, and finally, 
(iii) the winds of LBVs can be used to derive their 
masses, as exemplified by the case of AG\,Car, for which we derive a 
present-day mass of 35 \msun.
}

\maketitle


\section{Introduction}
\label{s_intro}

Luminous Blue Variables (LBVs) are massive stars in a brief
($\sim$\,10$^{5}$ yr) but violent post-main-sequence phase of 
evolution, during which they lose the remaining parts of their 
hydrogen-rich outer layers, before entering the Wolf-Rayet stage. 
They show a complex pattern of visual brightness variations (Humphreys 
\& Davidson 1994; Lamers 1995; van Genderen 2001), but most characteristic are
their visual light variations of $\Delta V$\,$\simeq$\,1 -- 2 magnitudes 
on timescales of years. These changes are referred to as ``typical'' or 
``S\,Doradus-type'' outbursts.
During these variability cycles, the stars move horizontally in the 
Hertzsprung-Russell Diagram (HRD) as they expand in radius at 
approximately constant luminosity.
At minimum visual brightness the star is relatively hot
($\teff$\,$\sim$\,20\,000 -- 30\,000 K) and small, whereas the star is large and 
relatively cool ($\teff$\,$\sim$\,8\,000 K) at visual maximum. 
Although it has been established that the typical variations are related 
to changes in stellar radius rather than the formation
of a pseudo-photosphere (Leitherer et al. 1989; de Koter et al. 1996), the 
physical reason for these radius changes -- which can be as 
large as factors of ten -- is as yet unidentified (see Nota \& Lamers 1997
for overviews).

The strong stellar winds of LBV stars 
($\mdot$\,$\sim$\,$10^{-4}$ -- 10$^{-6}$ \msunyr) show a wide variety of mass loss 
behaviour, which is poorly understood.
There are cases 
like R\,71 where the mass loss increases while the star expands 
(Leitherer et al. 1989), whereas R\,110 shows a behaviour 
that is the exact opposite: while the star increases in size, 
its mass-loss rate drops (Stahl et al. 1990). These  
results imply that there is no straightforward correlation 
between mass loss and radius (or effective 
temperature) for LBVs (Leitherer 1997).

Recent radiation-driven wind models for normal O and B stars 
however show that one does not expect such a strict monotonous 
behaviour of mass loss with effective temperature. Vink 
et al. (1999, 2000) have found an overall trend 
that cooler stars -- at constant luminosity -- experience smaller 
mass loss, but at specific effective temperatures the mass-loss 
rate jumps upward. 
The overall trend can be explained by a growing
mismatch between the position of the bulk of the 
driving lines (mostly in the ultraviolet) and the location of the 
flux maximum, which shifts gradually towards the optical 
for cooler stars. Superimposed on this there are jumps,
where \mdot\ increases steeply due to recombinations of dominant
line-driving ions, specifically those of iron. At 
$\teff$\,$\simeq$\,25\,000 K, the mass loss
is predicted to increase by about a factor of five due to the
recombination of Fe\,{\sc iv} to Fe\,{\sc iii}. Observational
evidence for this so-called `bi-stability' jump has been presented
by Lamers et al. (1995), who discovered an abrupt discontinuity
in the carbon ionisation and the terminal wind velocities of
stars at spectral type B1 ($\teff$\,$\simeq$\,21\,000 K). They also 
present evidence for a second jump at spectral type A0
($\teff$\,$\simeq$\,10\,000 K), which is
anticipated to be due to the recombination of Fe\,{\sc iii} to
Fe\,{\sc ii} (Leitherer et al. 1989).

The main goal of this paper is to investigate whether radiation
pressure on spectral lines is responsible for the mass-loss rates and 
mass-loss variability of LBV winds, or that alternative mechanisms, such 
as pulsations or magnetic fields, need to be invoked. 
If radiation-pressure is the main driver, just as it is 
in OB-type stars, then one expects the effective 
temperatures at the epochs 
at which $\mdot$ has been deduced, to determine whether an increase 
or decrease in mass loss occurs. 
In such a picture, the seemingly contradictory  
mass loss behaviour of e.g. R\,71 and R\,110 may be explained 
in terms of their differing effective temperature intervals with respect 
to the bi-stability jumps (de Koter 1997; Lamers 1997).

The mass-loss rates of LBVs are expected to differ from
those of normal OB stars -- at the same position in the 
HRD -- as the result of two effects.
First, their stellar masses (and consequently their mass over
luminosity ratios) are expected to be lower as LBVs are more
evolved. Second, their surface abundances are expected to be 
different as products of nucleosynthesis may have reached 
the stellar surface. Observations show that 
LBVs are indeed enriched in helium and nitrogen, and depleted in carbon 
and oxygen (e.g. Smith et al. 1994; Najarro et al. 1997). 
The LBV mass-loss predictions presented in this paper take these
effects into account.

As LBVs have the property that their effective 
temperatures change substantially on timescales of years, they offer
the unique opportunity to study mass loss as a function of effective 
temperature for individual stars. In this respect it is interesting to 
note that Stahl et al. (2001) have recently observed and 
analysed the Galactic LBV AG\,Carinae and found an 
increase in the mass-loss rate by a factor of five during 
its excursion from minimum to maximum visual light. 
Intriguingly, this increment in mass loss is similar in magnitude 
to the predictions of OB star mass loss around the bi-stability 
jump (Vink et al. 1999).

The paper is organised as follows. We first describe the Monte Carlo 
method used to predict mass-loss rates (Sect.\,\ref{s_method}). 
In Sect.\,\ref{s_massloss} we present the LBV mass-loss predictions. 
A comparison between these theoretical 
mass-loss rates and all available mass-loss observations 
is subsequently presented in Sect.\,\ref{s_obs}. Special emphasis 
is attributed to AG\,Car (Sect.\,\ref{s_agcar}), which is by far 
the best studied member of its group. Interestingly, the set 
of \mdot-determinations for this star provides such tight 
constraints that, in addition, the stellar mass of AG\,Car 
can be determined (Sect.\,\ref{s_mass_agcar}). We present our 
conclusions in Sect.\,\ref{s_concl}.


\section{Method to predict mass loss}
\label{s_method}

\subsection{A brief description of the Monte Carlo method}

We use Monte Carlo calculations of photons travelling through
an atmosphere that accounts for an outflowing wind to predict 
mass-loss rates. In essence, the mass loss follows from equating
the gain of kinetic energy of the outflowing gas to the decrease
of radiative energy. From an outside observer's point of view, this 
decrease of radiative energy occurs when photons transfer 
momentum (and energy) in interactions with moving ions
in the flow. An iterative process is used to find the model in
which the input mass loss is equal to that found from the global
radiative to kinetic energy conversion. This \mdot\ value is
the predicted mass-loss rate. Details of the Monte-Carlo method are 
given in Abbott \& Lucy (1985), de Koter et al. (1997) and Vink 
et al. (1999). The model atmospheres used in the process 
are calculated using the non-LTE code {\sc isa-wind} and involve a 
gradual transition of density, consequently of velocity (through 
the equation of mass continuity), and temperature between the stellar 
photosphere and wind region. The chemical species for which the
statistical equilibrium equations are solved explicitly are
H, He, C, N, O, and Si. For a full description of the code, 
we refer the reader to de Koter et al. (1993, 1997). The iron-group 
elements, which are important for the radiative driving, are not
explicitly accounted for but are treated in a generalised version
of the ``modified nebular approximation'' described by Lucy (1987, 1999).

Our method to predict mass loss is distinct from the approach
in which the line force is parameterised in 
terms of force multiplier parameters (e.g. Castor, 
Abbott \& Klein 1975, Pauldrach et al. 1994). The main advantage
of our approach is that it accounts for multiple photon scattering
processes, which have been found to be important even for normal
O- and B-type stars (e.g. Vink et al. 2000), but which are not
treated in the line-force-parameterisation method. 
A disadvantage of our approach may be that we do not {\it de facto} 
solve the momentum equation, i.e. the 
predicted mass loss values depend to some extend on the 
adopted velocity stratification $v(r)$. For this we assume a 
$\beta$-type velocity law, which has been found to yield excellent 
results in modelling the spectra of hot star winds (e.g. Puls 
et al. 1996). We estimate the intrinsic accuracy of our 
Monte Carlo calculations to be $\la$ 0.05 dex in the predicted \mdot. 

\subsection{Assumptions in the models}
\label{s_model_assumptions}

\paragraph{Modified nebular approximation.~} 

The simplified treatment of the ionisation and excitation state
of the iron-group elements, i.e. by means of the modified
nebular approximation, may result in a (systematic) shift of
the temperature at which the dominant ionisation state of
these species changes relative to full non-LTE calculations. 
An indication that such an offset may indeed be 
present has been presented by Vink et al. (1999). These 
authors identify the occurrence of strong changes in the terminal 
velocity and ionisation of the winds of OB supergiants at spectral 
type B1 (i.e. at $\sim$\,21\,000 K; Lamers et al. 1995) to be a result 
of the transition of the dominant iron ionisation from Fe\,{\sc iv} to 
{\sc iii}. Using a similar approach as the one used in this
study, they predict a jump in the mass-loss rate as a result of 
this recombination, but they predict the location of this
jump to be at $\sim$\,25\,000 K, i.e. at an effective temperature of $\sim$\,4\,000 K 
higher than the observed temperature of the jump. 
OB supergiant simulations over a wider range of wind densities indicate that  
the ionisation balance of Fe {\sc iv}/{\sc iii} is 
mostly sensitive to temperature, and not density. Therefore, the bi-stability 
jump was found to occur over only a relatively narrow range of $22.5 \leq \teff \leq 26$ kK (Vink 
et al. 2000). 
In other words, the offset between the predicted bi-stability jumps 
and the observed B1 jump was found to be in the range 1\,500\,--\,5\,500 K, with 4\,000 K being 
a typical value. 
One may therefore expect that for the dense winds of
LBVs, the offset will be at least of the same magnitude 
as the maximum B supergiant offset, i.e. 5\,500 K. 
We will later see (in Sect.\,\ref{s_agcar}) that the offset  
is $\simeq$ 6\,000 K.
We note that for stars that vary their effective temperatures in a range 
without iron recombinations, one does not expect the modified nebular approximation 
to affect the predicted mass loss. Indeed, in the case of O-type stars  
(where Fe\,{\sc iv} is dominant over the full O-star 
temperature range) good agreement was found between our predictions 
that use the modified nebular approximations and the observed mass-loss rates 
(Vink et al. 2000; Benaglia et al. 2001). 

In view of the above, the use of the modified nebular approximation  
is anticipated to cause an offset in the predicted position of the 
bi-stability jumps. We will apply a corrective shift to the locations 
of the jumps in our models to facilitate 
a meaningful comparison between these predictions and the observations 
(Sect.\,\ref{s_obs}).

\paragraph{Stationarity.~}

We assume a stationary stellar wind, i.e. both the mass-loss rate and
the velocity stratification are time-independent. The assumption of 
stationarity may not be too bad an assumption for predicting the 
time-averaged mass-loss rates of O stars (see e.g. Owocki 
et al. 1988). However, for LBVs this may not be fully correct as the 
timescale of photometric variability $\tau_{\rm phot}$ becomes 
comparable to, or shorter than, the typical 
flow timescale of the wind. If we define the latter as the timescale 
$\tau_{\rm dyn}$ for a gas element to travel from the photosphere to 
100 stellar radii -- typically the extent of the H$\alpha$ formation
region in LBVs -- we find at minimum visual brightness $\tau_{\rm dyn}$\,$\sim$ 
days, while at maximum visual light $\tau_{\rm dyn}$\,$\sim$ months. 
For LBVs, significant photometric changes may occur on timescales of
$\tau_{\rm phot}$\,$\sim$ months, and therefore the assumption of a stationary
wind is first expected to break down for LBVs in their transition from maximum
to minimum visual light. 
One should realise that such a breakdown would not only occur in  
wind models to predict mass loss, but it would also affect the analysis 
of LBV spectral observations, e.g. the modelling of the H$\alpha$ line 
profile in order to determine the mass-loss rate. A stationary
model atmosphere cannot predict the line cores (which are formed at low 
projected velocities) and the line wings (formed at high velocities) at the same 
time, as this would introduce significant errors in the deduced \mdot\ values. 

For these reasons we focus our comparison of predictions with
observations on the transition of LBVs from minimum to maximum visual light, 
and we do not include the reverse trajectory (see Sect.\,\ref{s_agcar}).

\paragraph{Sphericity.~}

Spectropolarimetric observations of the LBV AG\,Car and 
other LBVs have shown that the winds of these stars are not spherically 
symmetric (see e.g. Schulte-Ladbeck et al. 1994). This may 
have implications for both our mass-loss predictions, as well as 
the modelling of the H$\alpha$ lines to deduce the observed 
mass-loss rates. As there is little quantitative information on 
the density contrast between the pole and equator in LBV winds, 
it remains to be seen how important these non-sphericity effects are 
with respect to the global wind properties.  
In any case there is no particular reason why a possible systematic 
non-sphericity effect would appear more prominent in one visual 
brightness state than in the other.

\subsection{The adopted parameters}
\label{s_assumptions}

\begin{table}
\caption[]{Adopted stellar parameters for the grid of LBV models
           presented in this study.}
\label{t_params}
\begin{tabular}{cccc}
\hline
\noalign{\smallskip}
log$L_*$ & $M_*$ & $\Gamma_{\rm e}$ & $\teff$\\
\noalign{\smallskip}
($\lsun$) & ($\Msun$) & & (kK) \\
\hline
5.5 &  30 -- 10  & 0.18 -- 0.53  & 10.5 -- 30.0\\
6.0 &  35 -- 25  & 0.48 -- 0.67  & 10.2 -- 30.0\\
\noalign{\smallskip}
\hline
\end{tabular}
\end{table}

\paragraph{Luminosities and Masses.~}
We calculate mass-loss rates for two series of LBV models, one
having a luminosity \logll\ = 5.5 and the other 6.0. The luminosities of
most LBVs fall in this range, e.g. R\,71 and R\,110 are representative
for the low luminosity value, while AG\,Car and R\,127 are
representative for the higher one. 
Since the masses of LBVs are poorly known, we adopt 
a range of masses for both series (see Table~\ref{t_params}).
For the series with \logll\ = 5.5, mass loss 
was computed for $M$ = 30, 25, 20, 15, 12 and 10 $\msun$. For the 
series with \logll\ = 6.0, the adopted masses were $M$ = 35, 30, and 25 
$\msun$.

The luminosity-to-mass ratio ($L/M$) can be expressed in terms
of the Eddington factor $\Gamma_{\rm e}$.
This factor is defined as the ratio between the Newtonian acceleration 
and the radiative acceleration due to electron scattering, and is given by
\begin{equation}
\label{eq_gammae}
\Gamma_{\rm e}~=~\frac{L \sigma_e}{4 \pi c G M}~=~7.66~10^{-5} \sigma_{e} 
\left(\frac{L}{\lsun}\right) \left(\frac{M}{\msun}\right)^{-1}
\end{equation}
where the constants have their usual meaning and $\sigma_{\rm e}$ is the 
electron scattering cross-section per unit mass. 
The value for $\sigma_{\rm e}$ depends on temperature and chemical 
composition.
The range in $\Gamma_{\rm e}$ relevant for our model grid is given in 
Col. 3 of Table~\ref{t_params}.
Note that $\Gamma_{\rm e}$ is considerably smaller than unity, signifying
that the classical Eddington limit is not exceeded. The reason for this 
relatively low $\Gamma_{\rm e}$ is that although the 
masses are lower than for normal OB supergiants, the value 
for the electron scattering cross-section $\sigma_{\rm e}$ is also lower. 
This is a result of helium enrichment in LBV atmospheres.

\paragraph{Effective Temperatures.~}
Models have input temperatures computed for effective temperatures 
between 11\,000 and 
30\,000 K with a stepsize of 2\,500 K in the range 12\,500 - 30\,000 K. 
This input temperature $T_{\rm in}$ is defined by the relation:
\begin{equation}
L = 4 \pi R_{\rm in}^2 \sigma T^4_{\rm in}
\label{eq_Tin}
\end{equation}
where $R_{\rm in}$ is the inner boundary of the atmospheric model. 
$R_{\rm in}$ is typically located at Rosseland optical depth 
$20 \la \tau_{\rm R} \la 25$. Note that the actual effective temperature 
\teff, which follows from the model computation, will be somewhat lower.
The effective temperature is defined at the point where the thermalisation 
depth measured in the center of the $V$ band (at 5555 \AA) equals 
$1/\sqrt{3}$ (see Schmutz et al. (1990) and de Koter et al. (1996) for a 
detailed discussion). 
For stars with relatively low mass loss and consequently mostly optically 
thin winds, such as the winds from normal OB supergiants, there is no  
difference between the input $T_{\rm in}$ and \teff.
But in the case of LBVs, which may lose mass at rates of up to $\mdot
\sim 10^{-4} \msunyr$, there is an offset between the two temperature
values, but this offset is relatively small. Note that the 
discrepancy between $T_{\rm in}$ and $\teff$ is more dramatic 
for Wolf-Rayet stars, which have much denser winds.

\begin{table}
\caption[]{Solar versus adopted LBV abundances. The latter show surface helium 
           enrichment and CNO processed material.}
\label{t_abund}
\begin{tabular}{cll}
\hline
\noalign{\smallskip}
Element & Solar            & LBV             \\
        & (mass fraction)  & (mass fraction) \\
\noalign{\smallskip}
\hline
H   & 0.68           &  0.38           \\
He  & 0.30           &  0.60           \\ 
C   & 2.9\,10$^{-2}$ &  2.9\,10$^{-4}$ \\  
N   & 9.5\,10$^{-4}$ &  9.5\,10$^{-3}$ \\ 
O   & 7.7\,10$^{-3}$ &  7.7\,10$^{-4}$ \\ 
\noalign{\smallskip}
\hline
\end{tabular}
\end{table}

\paragraph{Abundances.~}
Both spectroscopic analyses and models for stellar evolution show
LBVs to be enriched in helium (Smith et al. 1994; Meynet et al. 1994; 
Najarro et al. 1997). In addition to helium enrichment, CNO processed 
material that has reached the surface causes an 
increase in the amount of nitrogen. This occurs at the cost of oxygen and 
carbon, which become depleted. 
The {\it total} amount of CNO remains nevertheless constant. 
The relative changes in the elements CNO have been adopted from the 
abundance computations for massive stars with rotation induced mixing by 
Lamers et al. (2001) and have been applied to the solar CNO abundances 
given by Anders \& Grevesse (1989). 
Table~\ref{t_abund} presents both the initial solar abundances (Col. 2) 
as well as the abundances modified by the CNO process (Col. 3). 
The metal abundance $Z$ is kept constant at the solar value, 
for models computed with all species included.

\paragraph{Velocity Laws.~}
We have calculated $\dot{M}$ for wind models with a $\beta$-type 
velocity law for the supersonic part of the wind:
\begin{equation}
\label{eq_betalaw}
v(r) = \vinf \left(1-\frac{R_*}{r}\right)^\beta
\end{equation}
where \vinf\ is the terminal flow velocity.
Below the sonic point, a smooth transition from this
velocity structure is made to one that corresponds to the
photospheric density structure and adopted \mdot\ (see de Koter et al. 1997
and Vink et al. 1999 for details). A value of 
$\beta=1$ was assumed in all calculations. The predicted mass-loss 
rates are found to be
insensitive to the wind acceleration parameter when 
$\beta$ is in the range 0.7 -- 1.5 (Vink et al. 2000).
We specified the terminal velocity adopting for the ratio $\ratio$
the values 1.3, 2.0, and 2.6. 
Lamers et al. (1995) found that a ratio of 2.6 for Galactic supergiants
of spectral type earlier than B1; a ratio of 1.3 for 
B1 to A0 stars, and a tentative ratio of 0.7 for later spectral types.


\section{The predictions of mass loss of LBVs}
\label{s_massloss}

Differences in the wind properties of an LBV star and a normal O- or
B-type supergiant in the same location of the HRD are expected to be 
the result of differences in stellar mass (specifically
in $L/M$) and in surface chemical composition.
Regarding the latter, in LBVs helium may be enriched at the 
surface and CNO-cycle processed material may have appeared. This 
yields an increase in N and a decrease in C and O relative to the initial
abundances. We will now discuss the difference in mass-loss rate
between LBVs and OB-stars as a result of these effects.

\begin{figure*}
\centerline{\psfig{file=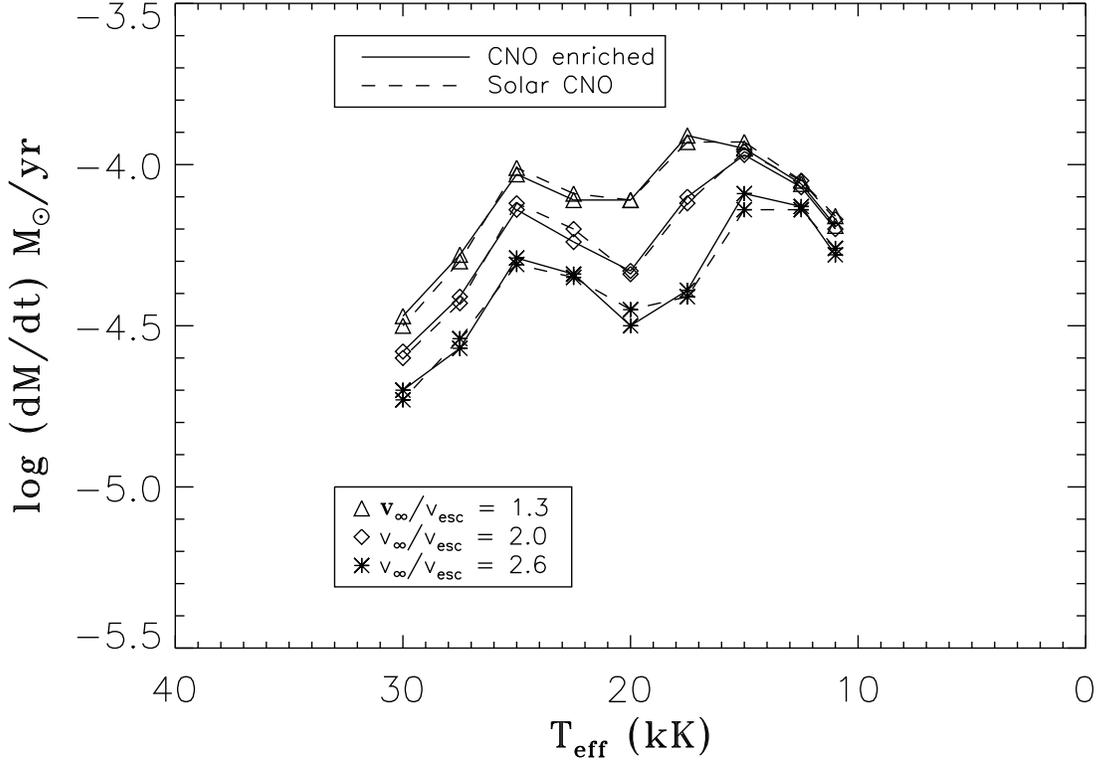, width = 16cm}}
\caption{Predictions of mass loss as a function of effective temperature
         for a star with solar (dashed lines) versus CNO processed
         surface abundances (solid lines). As CNO serves as a catalyst in the
         nuclear reactions the total number C+N+O atoms is invariant.
         The adopted stellar parameters are
         \logll = 6.0; $M = 30 \msun$; (X,Y) = (0.68,0.30). Adopted
         values for the ratio between terminal and effective escape 
         velocity are given in the legend.
        }
\label{f_cno}
\end{figure*}

\subsection{The effect of CNO processed material on \mdot}
\label{s_cno}

Figure~\ref{f_cno} shows a comparison in mass-loss behaviour between
models with initial (solid lines) and CNO processed abundances (dashed
lines). The results are very similar, i.e. the occurrence of CNO
processed material does not change the predicted mass loss in any
major way. There are two reasons for this. First, in the case of
Galactic abundances (which are being discussed here) the dominant 
wind driving element is iron, relative to which CNO only plays a modest 
role in determining the mass-loss rate (Vink et al. 1999, Puls et al. 
2000). Second, as the CNO elements only serve as a 
catalyst in the nuclear reactions, the total number C+N+O remains constant. 
The net effect of the nitrogen enrichment and the carbon/oxygen depletion on 
the line driving is therefore expected 
to be small, as it relies on differential effects in the number
of lines provided by the relevant ions of these species.

\begin{figure*}
\centerline{\psfig{file=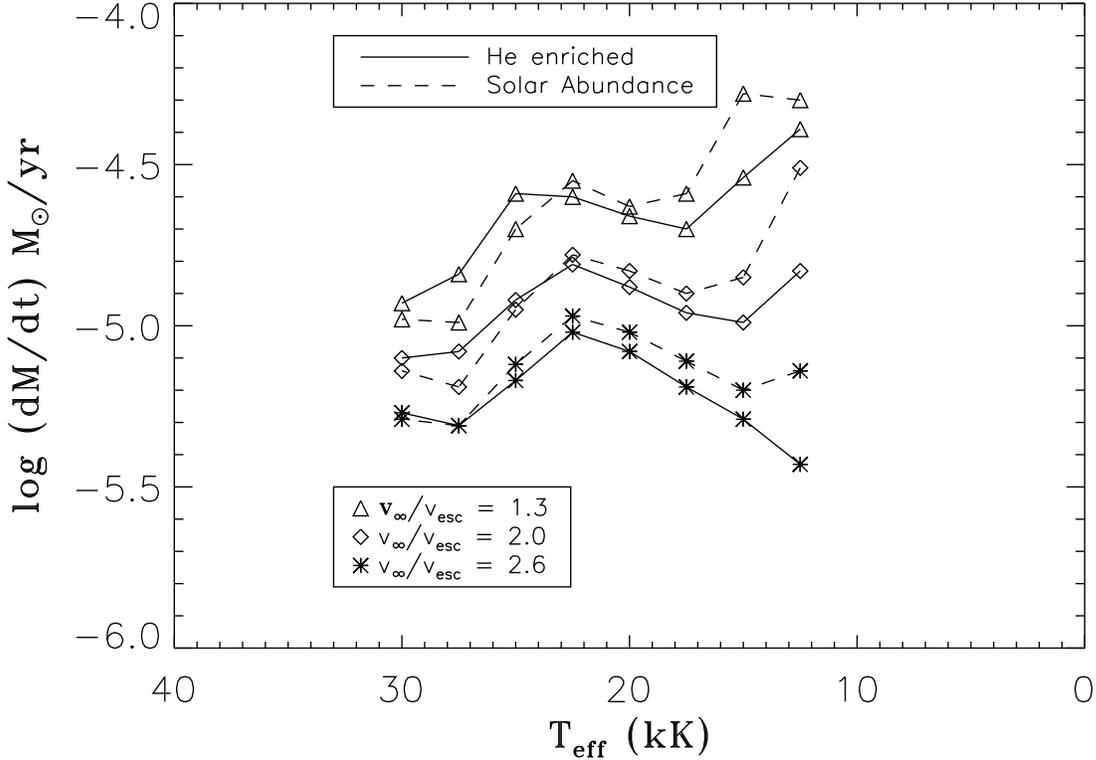, width = 16cm}}
\caption{Predictions of mass loss as a function of effective temperature
         for a star with solar (dashed lines) versus helium enriched
         (X=0.38, Y=0.60) abundances (solid lines). The adopted stellar
         parameters are $\log (L/\lsun)$ = 6.0 and $M = 60 \msun$.
         Adopted values for the ratio between terminal and effective
         escape velocity are given in the legend. 
        }
\label{f_helium}
\end{figure*}

\subsection{The effect of helium enrichment on \mdot}
\label{s_helium}

To investigate the difference in mass-loss behaviour as a result of
changes in the hydrogen and helium abundance, we compared predictions 
for (X,Y) = (0.68,\,0.30) and (0.38,\,0.60). The models 
have \logll = 6.0 and $M$ = 60 \msun, and are presented in 
Figure~\ref{f_helium}. The solar abundance models (dashed lines) have been 
taken from Vink et al. (2000). The models with enhanced helium (solid lines)
are characteristic for what is expected in the LBV phase (Meynet et al.
1994).

The most pronounced difference between the two sets of models is that
at $T \ga 25\,000$ K (for \ratio\ = 1.3 and 2.0) the helium enriched models 
show a {\em higher} mass loss, while at lower temperatures they
feature a {\em lower} mass loss. The difference in \mdot\
extends up to $\sim$\,0.2 dex. A second difference is that the position
of the bi-stability jump at $\sim$\,25\,000 K (where the iron ionisation
in the region of strong wind acceleration flips between Fe\,{\sc iv} and 
{\sc iii}) appears to shift slightly towards higher temperature, while
the opposite occurs for the jump at $\sim$\,15\,000 K (i.e. 
where Fe\,{\sc iii} changes over to Fe\,{\sc ii}). 
 
The cause of this behaviour is {\it not} the result of a 
difference in the line driving of H~{\sc i} relative 
to He~{\sc i} and He~{\sc ii}, as hydrogen 
and helium are both irrelevant for the line driving 
force at Galactic abundances (Vink et al. 2001).
As it is the {\it iron} lines that dominate 
the wind driving (below the sonic point), the relevant 
question concerns the dominant ionisation stage of iron, 
and how well the wavelength positions of the Fe lines match the 
spectral region where the bulk of the flux is emitted.
The many lines of Fe\,{\sc iv} and Fe\,{\sc iii} are however 
spread throughout the Lyman and Balmer continua, and do not 
show a particularly strong preference for either side of 
the Lyman jump (at 912 \AA). Therefore, the cause of the 
differences in mass loss behaviour must instead be related to 
changes in the underlying flux distribution, in particular 
the strength of the Lyman jump.

When hydrogen is (partly) replaced by helium, the size of the 
Lyman jump decreases. This implies that more photons are available 
in the Lyman continuum (at wavelengths below 912 \AA). Since the 
luminosity of the models is kept constant, less flux will be emitted
in the Balmer continuum. At relatively low \teff, the bulk of the
flux is emitted in the Balmer continuum, therefore one expects the
mass loss in the helium-enriched models to be lower; at relatively
high \teff, most of the flux is emitted in the Lyman continuum, and as 
the flux in the helium-enriched models is higher here, more 
mass loss occurs.

The small differences in the $\teff$ positions of the two bi-stability 
jumps can be explained in a similar way. In the regime of the Fe\,{\sc iv}
to {\sc iii} jump at $\sim$\,25\,000 K, the wind densities in the
helium-enriched models are higher, therefore recombination/ionisation
will occur at a slightly higher temperature relative to the solar abundance
models. For the Fe\,{\sc iii}/{\sc ii} jump at $\sim$\,15\,000 K the
situation is reversed: the wind densities in the helium-enriched models
are lower, therefore the jump occurs at somewhat lower \teff.

\subsection{The effect of lower stellar masses on \mdot}
\label{s_mass}

\begin{figure*}
\centerline{\psfig{file=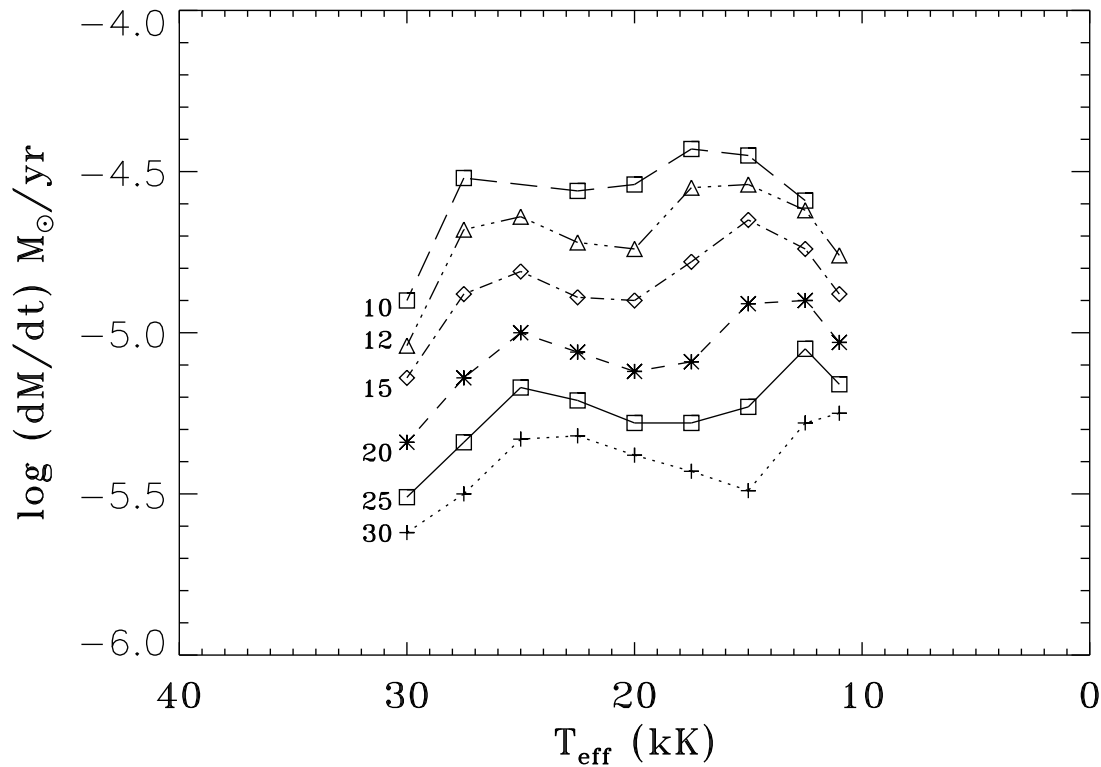, width = 16cm}}
\caption{Predictions of mass loss as a function of stellar mass for a
         star with solar abundances. The adopted luminosity is
         \logll\ = 5.5. We find that 
         $\log \mdot \propto -1.8 \,\log M$. The different masses in \msun\ are 
         given in the plot.}
\label{f_mass}
\end{figure*}

The effect of decreasing the stellar mass is shown in Fig.~\ref{f_mass}
for a set of models that have \logll\ = 5.5. The calculations
show that for all temperatures the mass loss increases with decreasing
stellar mass. For this luminosity, Vink et al. (2000) derived 
$\log \mdot \propto -1.3 \,\log M$ for masses in the range
between 30 and 50 \msun. Here, we find that the predicted dependence 
is given by 

\begin{equation}
\log \mdot~\propto~-1.81~(\pm \, 0.05)~\log M 
\label{eq_mass}
\end{equation}
for masses from 30 down to 10 \msun. 
For this mass range, this steep dependence implies an 
increase of the mass loss by a factor of seven, when the 
mass is decreased by a factor of three (from 30 to 10 \msun). 
This strong sensitivity can, in principle, 
be used as a diagnostic to derive current LBV masses. 
We will return to this in Sect.\,\ref{s_agcar}.

Also visible in Fig.~\ref{f_mass} is a gradual shift of the location
of the bi-stability jumps towards higher 
temperature for lower stellar masses. This is as one would expect: 
the models with low stellar mass have a relatively high mass loss,
and consequently high wind density. The Saha-Boltzmann ionisation 
equilibrium (as formulated in the modified nebular approximation) then 
implies that recombination will occur at a somewhat higher temperature.

\begin{figure}
\centerline{\psfig{file=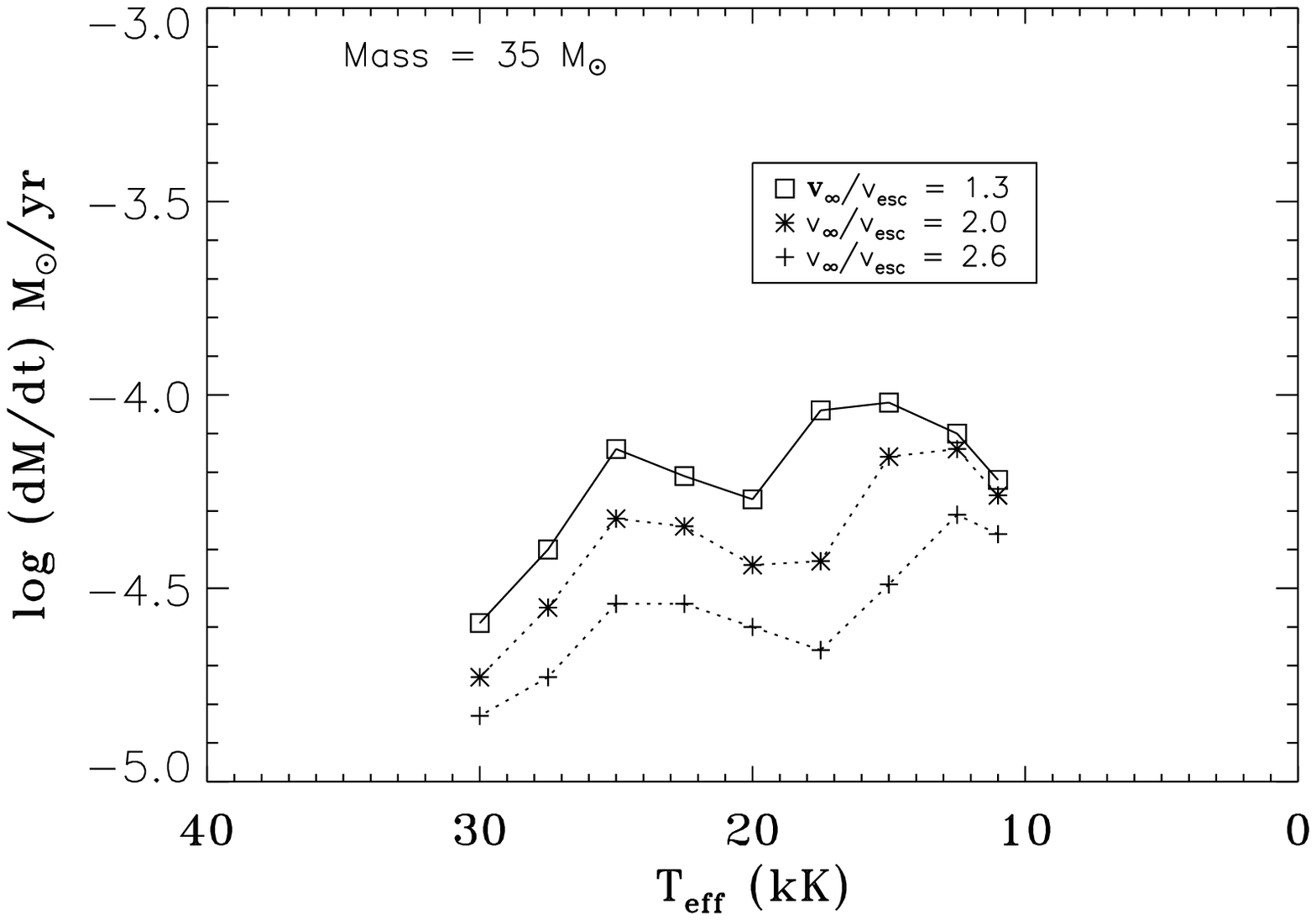, width = 9 cm}}
\centerline{\psfig{file=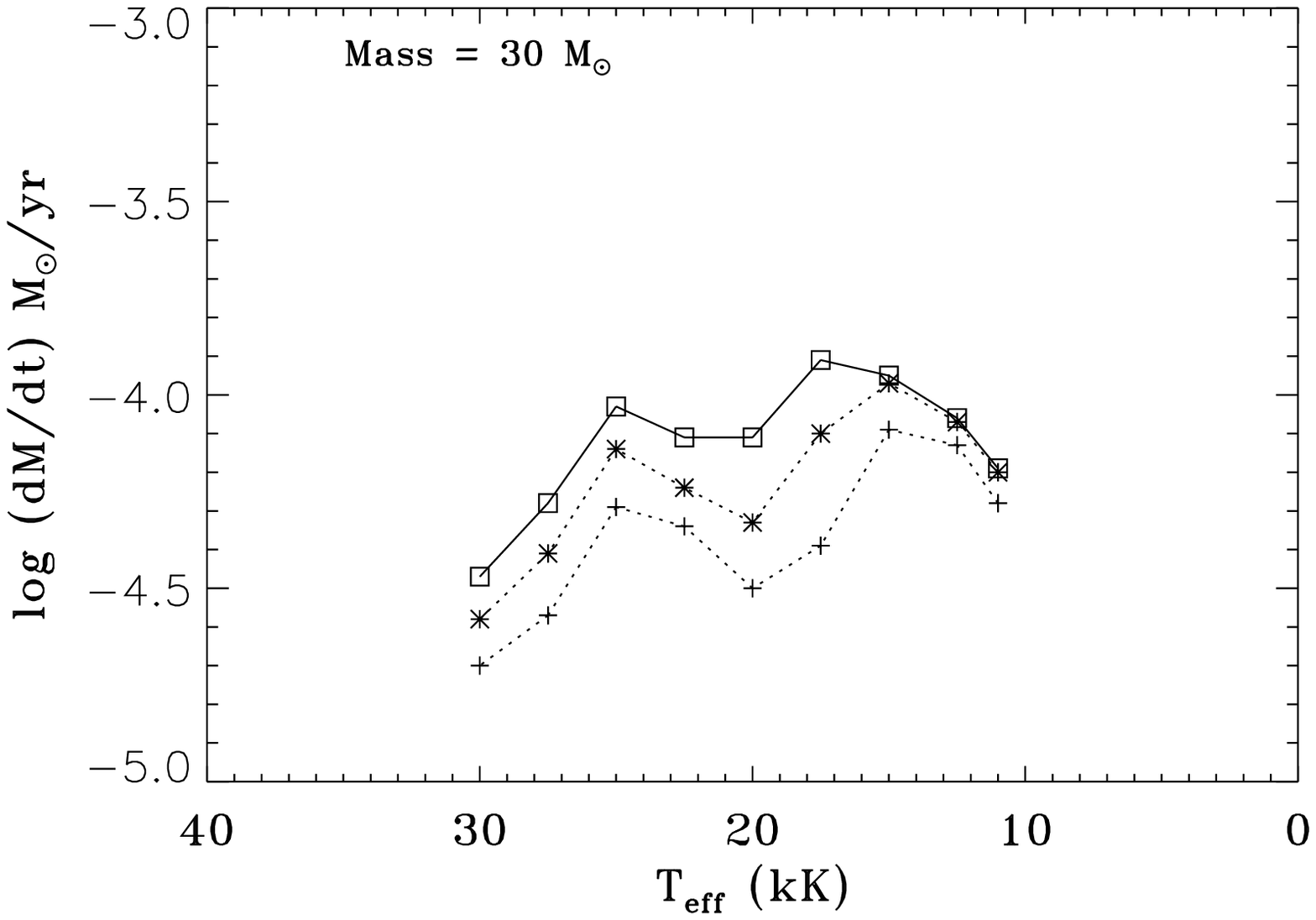, width = 9 cm}}
\centerline{\psfig{file=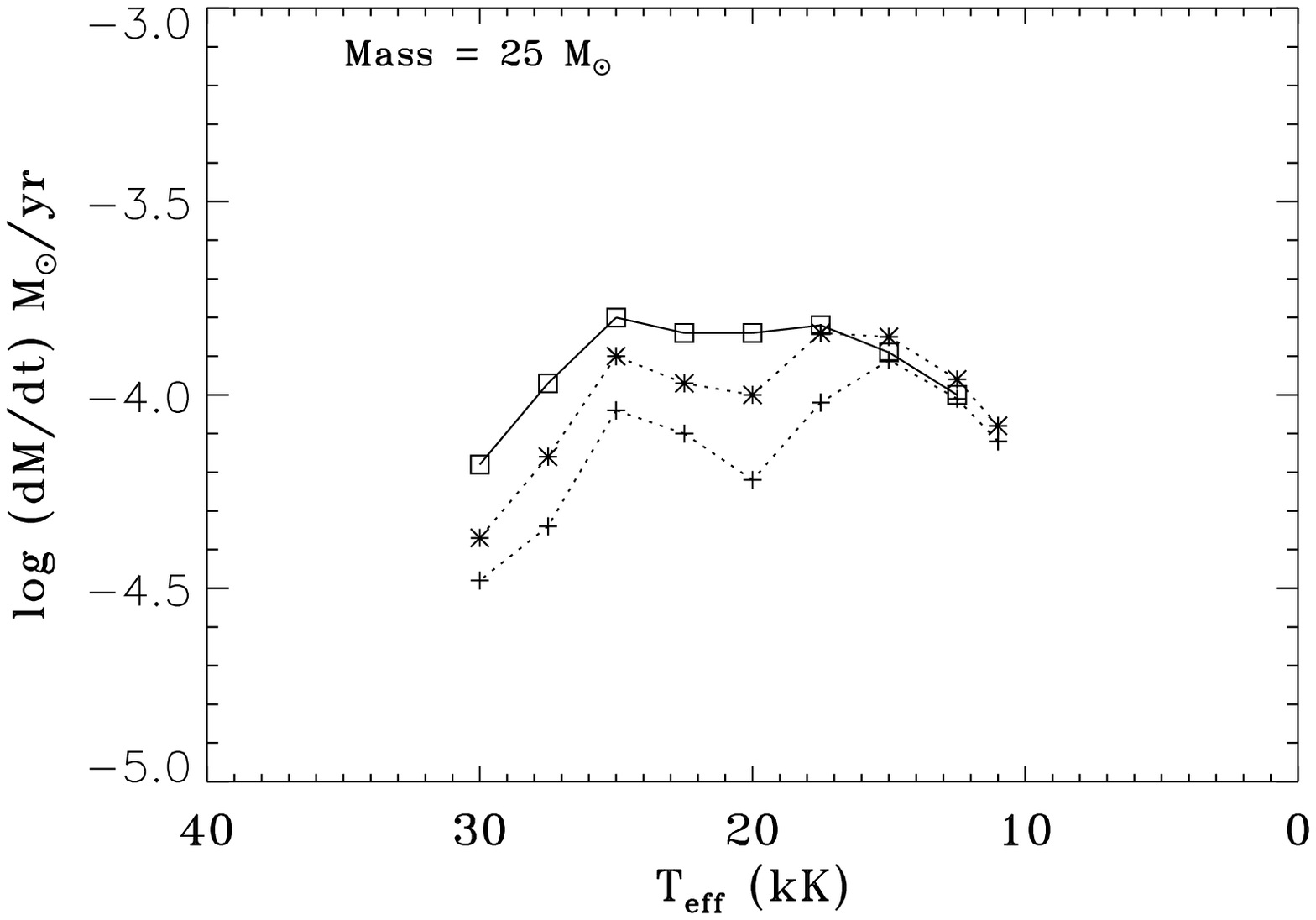, width = 9 cm}}
\caption{Predicted mass-loss rates as a function of effective temperature
         for three adopted stellar masses. All models have log (L/$\lsun$) 
         = 6.0. The values for the ratio between terminal and effective
         escape velocity are given in the upper panel.}
\label{f_mdots}
\end{figure}

For the series with luminosity \logll\ = 6.0 predictions for the mass-rate
as a function of stellar mass are presented in Fig.~\ref{f_mdots}. The
calculations for 60 and 80 \msun\ shown in Vink et al. (2000) have 
been extended here to 35 (top panel), 30 (middle panel), and 25 \msun\ (lower
panel). Results are shown for three different ratios of \ratio.
The same trends are visible and the same mass-loss vs. stellar mass 
dependence can be inferred.

\subsection{Summary:}
We conclude from the comparison between the mass loss of normal OB
supergiants and LBVs that:
(1) CNO processed material at the surface of LBVs does not result in a
    noticeable change in \mdot\ as the total number of effective driving 
    lines is not significantly affected.
(2) Helium enrichment in LBVs causes a modest change in mass loss 
    which can be traced back to a decreased Lyman jump at 912 \AA.
(3) The reduced stellar mass of LBVs causes a strong increase in the 
    mass-loss rate relative to normal OB supergiants.


\section{Comparison between LBV predictions and observations}
\label{s_obs}

\subsection{Observed mass-loss behaviour of LBVs}

In a recent census, van Genderen (2001) identifies a total number 
of 34 LBV members in the Galaxy and the Magellanic Clouds. 
He divides the LBVs into three groups: strong-active members, characterised 
by light amplitudes of $\Delta V > 0.5$ mag; weak-active members 
showing $\Delta V < 0.5$ mag; and ex- and dormant members which have not shown 
any typical photometric variability in the previous century. 
As our interest is in the mass loss behaviour of LBVs as a function of 
effective temperature, we are especially interested in the strong-active members. 
Although all of these 14 stars have been monitored photometrically over 
the last decades, only a handful have been subject to quantitative 
spectroscopic investigation at different epochs of photometric variability. 
This implies that estimates of the basic stellar parameters, such as the 
mass-loss rate, are rarely found at different epochs. 
In this respect, AG\,Car is by far the best investigated member, with 
detailed studies undertaken by Leitherer et al. (1994), Schmutz (1997) and most recently 
Stahl et al. (2001). We will focus on AG\,Car in Sect.\,\ref{s_agcar}, 
but we first present a brief overview of the other three LBVs that have been 
investigated spectroscopically. These LBVs are all in the Large 
Magellanic Cloud.

\paragraph{R\,71:}
Spectra of Radcliffe 71 at minimum and maximum visual light have been discussed 
in detail by Wolf et al. (1981). They derive temperatures
for the two phases of $\teff \sim 13\,600$ and 6\,000 K respectively. The
luminosity of the star is about 10$^{5.5}$ \lsun. From
a simple H$\beta$ line profile analysis they determine 
\mdot\ = $10^{-6.5}$ \msunyr\ for the hot phase. For the cool phase 
they estimate from the observed infrared excess that 
\mdot\ = $10^{-4.3}$ \msunyr. In a historical
context, it is essentially this difference of a factor 100 -- 200 that 
spurred the idea of explaining the radius changes of LBVs by a variable
mass loss, leading to the formation of pseudo-photospheres at maximum
light (Davidson 1987, Lamers 1987). However, for two reasons we suspect 
that the derived maximum mass loss may be incorrect. First, dust emission
contributes to the near-infrared flux of R\,71 (e.g. Voors et al. 1999) 
which Wolf et al. (1981) have not corrected for. This implies that they have 
overestimated \mdot. Second, the change in the H$\beta$ profile between
minimum and maximum light does not corroborate a change in mass loss
of a factor 100-200. In case H$\beta$ is optically 
thin, it holds that
 
\begin{equation}
\label{eq:mdotratio}
\frac{\mdot_{\rm max}}{\mdot_{\rm min}} \propto 
\left( 
\frac{\cal{F^{\rm c}}_{\rm max} \, W^{\rm eq}_{\rm max}}
     {\cal{F^{\rm c}}_{\rm min} \, W^{\rm eq}_{\rm min}}
\right)^{1/2}
\end{equation}
where ${\cal{F^{\rm c}}}$ denotes the continuum flux at 4861 \AA\ and 
$W^{\rm eq}$ represents the equivalent width of H$\beta$. 
The above equation assumes that the terminal velocity does not 
vary between minimum and maximum phase, which has indeed been found to 
be the case (Wolf et al. 1981). To estimate the mass loss ratio 
between visual minimum (min) and maximum (max) light, we use 
the H$\beta$ peak strength ratios, as given by Leitherer 
et al. (1989), of respectively 1.3 and 5.0 to approximate 
the equivalent width ratio. 
We estimate the continuum flux ratio using the total visual 
magnitude change of $\Delta V \sim 1$ mag. We then
arrive at a mass loss at maximum visual light of $\sim$\,3\,$\mdot_{\rm min}$. 
In case the line would be optically thick, and the power-law index 
in Eq.~(\ref{eq:mdotratio}) is alternatively given by 
4/3 (de Koter et al. 1996), 
we find that $\mdot_{\rm max} \sim$\,5.5\,$\mdot_{\rm min}$. 
Note that model calculations of Leitherer et al. (1989) also 
indicate that the change in the H$\beta$ profile points to a 
much more modest difference in mass loss between the two phases than 
the factor 100-200 estimated by Wolf et al. (1981).
We come to the conclusion that R\,71 shows an increase in mass loss
of a factor of 3 to 5 when it changes its temperature from
$\sim$\,13\,600 to 6\,000 K. Our models indeed predict a change of this
magnitude as the star crosses the Fe\,{\sc iii} to Fe\,{\sc ii} 
recombination jump at 10\,000 K.

\paragraph{R\,127:} R\,127 has been extensively monitored by the
Heidelberg group (Stahl et al. 1983, Stahl \& Wolf 1986, Wolf et al. 1988).
Based on spectral type determination, the stellar temperature is found to
vary between about 30\,000 and 10\,000 K. The authors note a rather 
capricious behaviour of the mass-loss rate. We anticipate such behaviour 
as two iron recombination jumps are located within this temperature 
range, i.e. Fe\,{\sc iv} to {\sc iii} at $\sim$\,21\,000 K and Fe\,{\sc iii} 
to {\sc ii} at $\sim$\,10\,000 K. Unfortunately, a lack of well determined 
mass-loss rates at different phases in its variability cycle prevents 
a quantitative comparison.

\paragraph{R\,110:} The lightcurve of R\,110 has been discussed in detail by
van Genderen et al. (1997). The star has been investigated
spectroscopically at two different epochs by Stahl et al. (1990). They
found that in the course of brightening, when the star changed its spectrum
from B-type with P\,Cygni lines to normal F0, the mass loss decreased from an
estimated few times 10$^{-6}$ \msunyr\ to a value that cannot be much 
higher than about 10$^{-6}$ \msunyr.
%
%
The implication of the finding that \mdot\ is smaller for
lower $\teff$ is that the cause of the excursions of LBVs in the
HR-diagram cannot be the formation of an opaque stellar wind as the
result of an increased mass loss. Our mass-loss predictions do indeed 
show $\teff$ intervals for which the mass loss increases inversely 
proportional to the temperature. The uncertainty in the derived 
mass loss by Stahl et al. (1990) is however too large to allow for a meaningful
quantitative comparison.

\begin{figure}
\centerline{\psfig{file=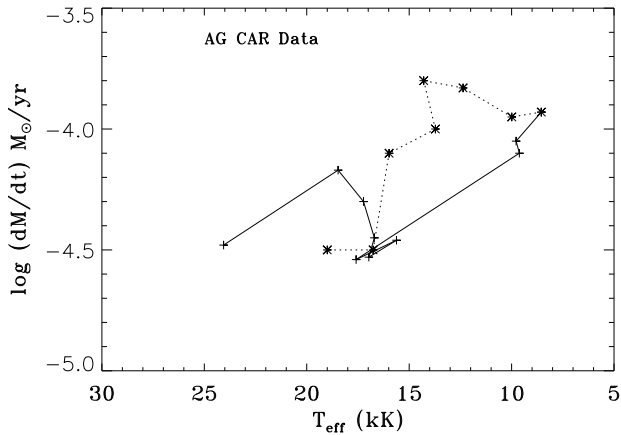, width = 9cm}}
\caption{Mass-loss rates of AG\,Car as a function of effective 
         temperature as derived from analysis of the H$\alpha$
         profile by Stahl et al. (2001). The solid line connects
         points on a rising branch in the lightcurve, i.e. when
         the star is on its way from minimum to maximum visual
         light, and covers the epoch Dec 1990 -- Feb 1995.
         The dashed line indicates the mass-loss behaviour on
         the falling branch, covering the period Feb 1995 -- Aug 1999.
        } 
\label{f_agcar}
\end{figure}

\subsection{The mass-loss behaviour of AG\,Car}
\label{s_agcar}

AG\,Car is by far the best observed LBV in terms of its mass-loss 
behaviour. Leitherer et al. (1994) derive \mdot(\teff) for the rising
branch in its lightcurve covering the period from December 1990 to
June 1992. Stahl et al. (2001) also investigate the mass loss behaviour
of the star covering the rising and falling branches in the epoch 
December 1990 -- August 1999. The methods in these papers to determine 
\mdot\ are to some extent rather similar. In both studies the luminosity
of the star is assumed to be constant at $\logll = 6.0$, and the temperature
is derived by fitting the observed flux in the $V$-band to that of
non-LTE model atmospheres (see the respective papers for details). 
The two methods differ in the way the mass loss is derived. Leitherer
et al. used the equivalent width of H$\alpha$, while Stahl et al.  
modelled the H$\alpha$ profile at selected phases. The latter method
is more accurate, and it yields more reliable $\mdot$ values; it additionally
allows one to set constraints on the velocity law. 
We will therefore compare our predictions to the data of Stahl et al. (2001).
Their derived mass-loss rates are plotted against effective temperature
in Fig.~\ref{f_agcar} for both the rising (solid line) and the
falling branch (dotted line). Note that there is a difference in 
behaviour between the star crossing over from minimum to 
maximum visual light, and visa versa. Our models do not predict such
behaviour, as the basic stellar parameters (including
stellar mass) are identical at similar positions in the HRD
on the rising and the falling branch.
We suspect that the differences in \mdot\ between the two branches, 
as derived by Stahl et al. (2001), are the result of the breakdown of 
the assumption of a stationary wind (see Sect.\,\ref{s_model_assumptions}) 
for the falling branch.
The long dynamical timescale of the flow at maximum light implies
that material lost in that phase may still be in close vicinity 
to the star, affecting the mass loss determination in the falling 
branch leading to a spuriously high (surface) mass-loss rate. 
A second explanation may be the release of gravitational energy 
when the star returns from maximum to minimum. If this plays 
a role, the assumption of constant bolometric magnitude during 
the variability is no longer valid (see Lamers et al. 1995). 
Whether or not AG\,Car remains at constant luminosity during 
its typical variations has however not yet been firmly established. 
Because of the two above-mentioned complications, we will focus
the comparison of our predictions to the observed mass-loss rates 
to the rising branch only.

\begin{figure*}
\centerline{\psfig{file=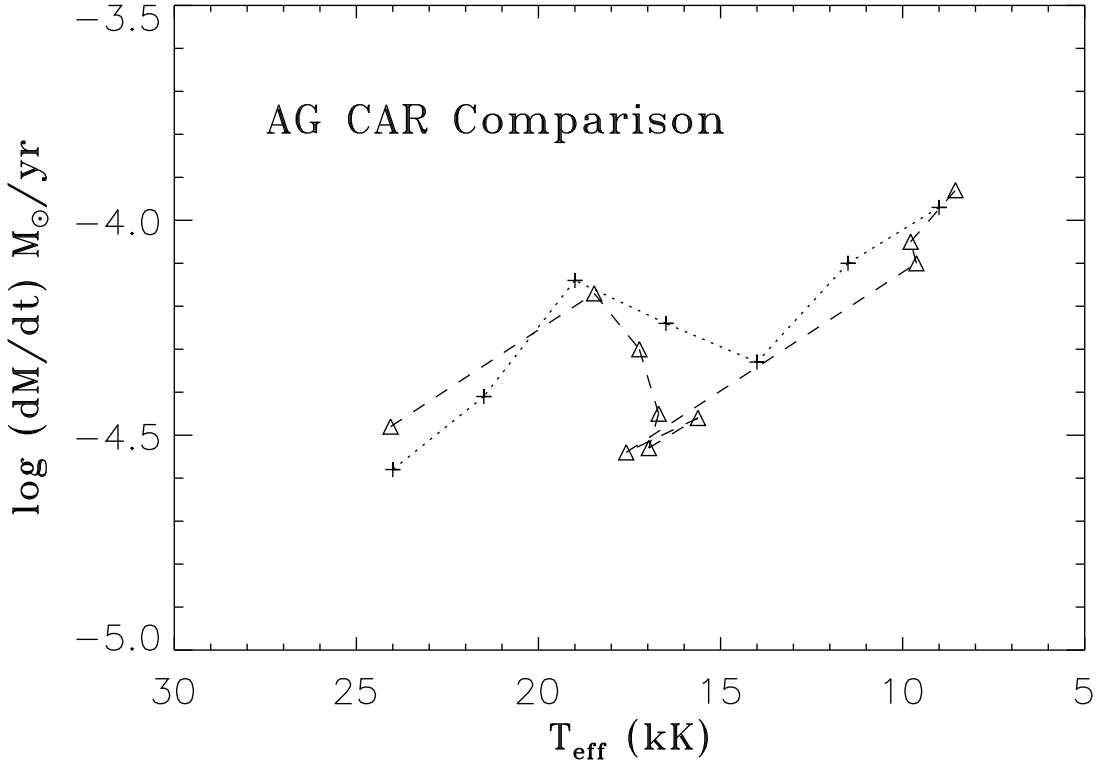, width = 16cm}}
\caption{Comparison of the predicted (dotted line) and 
         observed (dashed line) mass loss of AG\,Car. The 
         adopted stellar parameters are \logll = 6.0; 
         (X,Y) = (0.28,0.60), and $\ratio = 1.3$. All calculations
         have been shifted by $\Delta \teff = -6\,000$ K to correct for
         a systematic offset in the predicted ionisation balance of
         iron. The good agreement allows us to determine 
         the present-day mass of AG\,Car to be 35 \msun\ 
         by fitting the mass-loss rates.}
\label{f_comp}
\end{figure*}

In Fig.~\ref{f_comp} we compare our mass-loss predictions for AG\,Car
with the rates determined by Stahl et al. (2001) for the rising branch over
the epoch Dec 1990 -- Aug 1999, where the temperature decreases from
24\,000 to 9\,000 K. In our calculations we adopt
   $\logll = 6.0$; 
   a stellar mass $M = 35 \,\msun$; 
   a helium mass fraction Y = 0.60, 
   and a constant ratio of terminal over escape velocity \ratio\ = 1.3. 
The luminosity and helium abundance are very similar to that assumed by 
Stahl et al. (2001). The stellar mass is a free
parameter or, phrased differently, fitting the mass loss
allows one to determine the mass (see Sect.\,\ref{s_mass_agcar}). 
Unfortunately, the ratio \ratio\ turns out to be rather poorly constrained
by observations, as H$\alpha$ only allows to determine a lower
limit to \vinf\ (Stahl et al. 2001). Taking these lower limits and
correcting the escape velocity for electron scattering, Stahl et al.  
find lower limits in the range $0.2 \la \ratio \la 1.1$. We 
have opted to use our lowest calculated \ratio\ for the whole
temperature range, as this is the most appropriate value. 
We need to make two remarks with respect to this choice.

First, the adopted ratio of the terminal over the effective escape 
velocity influences the present-day mass determination of the star. 
For instance, if we would adopt a value \ratio\ = 2.0 we 
would find $M = 30\,\msun$, while a choice of \ratio\ = 2.6 would yield 
$M = 25\,\msun$. To first order, this positive relation between 
the velocity ratio and the stellar mass is explained as follows. For 
fixed stellar parameters, the photon momentum ($L/c$) transfered to 
the wind ($\mdot \vinf$) is more or less constant. It would therefore be  
necessary to decrease the stellar mass to match the observed mass-loss 
rates. However, if the actual value of \ratio\ were less than 1.3, we 
would anticipate that the mass would be {\em lower} than 35 \msun\, rather 
than higher. This because Doppler shifting of lines is strongly reduced 
for low velocity winds, which decreases the effectiveness of wind driving 
and would indeed require a lower stellar mass to match the observed \mdot.

Second, the temperature range investigated by Stahl et al. (2001) 
includes the regime where iron recombines from Fe\,{\sc iii} 
to Fe\,{\sc ii}. As a result of this, we predict a jump in mass 
loss around 15\,000 K. Possibly, this jump is related to the drop 
in the ratio \ratio\ from about 1.3 to 0.7 at spectral 
type A0, as identified by Lamers et al. (1995) on the basis of 
observed terminal velocities of supergiants. Because A0\,I stars 
have $\teff$\,$\sim$\,10\,000 K, a similar shift between predicted and 
observed ionisation appears to be present as was found for 
the recombination of Fe\,{\sc iv} to Fe\,{\sc iii} around spectral type B1 
(cf. Sect.\,\ref{s_model_assumptions}). 
In order to compensate for this offset, we shift our
calculations by $\Delta \teff = -6\,000$ K. The observed change in
terminal velocity at spectral type A0 would affect the predicted 
\mdot\ for the points at visual maximum light (i.e. at 
$\teff$\,$\la$\,10\,000 K). As this is at the very low end of 
the investigated 
temperature range, we have opted not to decrease \ratio\ further, i.e. 
below 1.3. 

Figure~\ref{f_comp} shows that after accounting for the corrective 
shift $\Delta \teff$, the observed and predicted mass loss agree 
within about 0.1 dex. Taking into account that \mdot(\teff) shows 
a complex behaviour, fluctuating over more than 0.5 dex, this 
is a surprisingly good result. It shows that
the mass loss variability of AG\,Car is the result of changes in
the ionisation of the dominant line-driving element, iron.

\subsection{The stellar mass of AG\,Car}
\label{s_mass_agcar}

The \mdot(\teff) behaviour can also be used to 
constrain the stellar mass.
The reason is that the predicted mass-loss rates are not only 
a function of \teff, but also of stellar mass (see Eq.\,\ref{eq_mass}). 
Comparing the AG\,Car data with \mdot(\teff) predictions 
for different masses results in a best fit for the present-day mass of 
35\,$\pm$\,5\,$\msun$. Note that this mass estimate does not account for  
possible systematic errors. Notably, one should be aware that the 
H$\alpha$ spectral modelling procedure of Stahl et al. (2001) did not account for clumping. 
In the case of Wolf-Rayet stars it is well-established 
that the neglect of a clumping correction may lead to systematic errors 
in the mass-loss rates by factors of $\sim$ 2\,--\,4 (see e.g. Hillier 2000). 
If clumping also plays an important role in the winds of LBVs, 
this would imply that the mass-loss rates derived by Stahl et al. 
are too high. Consequently, our mass derivation would be underestimated. 
We will now discuss whether a present-day mass of 35\,\msun\ is a reasonable value for AG\,Car.

Applying multiple distance indicators, Humphreys et al. (1989) constrain 
the distance to AG\,Car to be 5 -- 7 kpc, implying a luminosity of 
$\logll = 6.22 \pm 0.16$. Comparison with evolutionary tracks 
(Schaller et al. 1992) indicates that the most appropriate initial mass 
for such a large luminosity equals 120 \msun\ (but it would be only 
85 \msun\ when main sequence mass loss is taken twice the canonical value;  
Meynet et al. 1994). The extreme brightness of such a star would imply 
a very high mass loss, which affects the evolution in a profound way. 
The evolutionary tracks show that a star with an initial mass of 120 \msun\ 
evolves to the Wolf-Rayet phase almost directly from the main sequence, without 
even entering a post-main sequence supergiant phase. This scenario is 
very difficult to reconcile with the (range of) spectroscopic properties 
observed for AG\,Car, which indicates that its luminosity must be less than 
$\logll = 6.22$. For these reasons, we adopt $\logll = 6.0$, a value 
which is still in fair agreement with the lower bound found by 
Humphreys et al. (1989). The corollary of all this 
is that the initial mass for AG\,Car is only 65 \msun, if we assume
that the star is indeed in a post-main-sequence phase.

Voors et al. (2000) have determined the mass of the dust shell around
AG\,Car and arrived at 0.25 \msun, with an accuracy of about
a factor of two. This error is dominated by the uncertainty in the
maximum size of the dust grains. Adopting the usual gas/dust ratio of 100, 
one finds a total circumstellar mass of about 25 \msun. 
If we assume that this material is of stellar origin, which is 
supported by the enhanced N abundance found in the ionised part 
of the nebula (Hutsem\'{e}kers \& van Drom 1991; Smith et al. 1998), the 
upper limit to the present-day mass of the star is $\sim$\,40\,\msun. 
This is in good agreement with the mass of 35 \msun\ that we 
have derived in this study.


\section{Summary and Conclusions}
\label{s_concl}

We have presented radiation-driven wind models for a set of stellar
masses and luminosities characteristic for LBVs
and have predicted their mass-loss rates. 
We studied the effects of lower masses and modified abundances on the mass-loss 
rates of LBVs in comparison to normal OB supergiants. 
We have found that the main difference is  
due to the lower masses of LBVs. In addition, we have found that an increased 
helium abundance changes the mass loss properties by small amounts 
(of up to about 0.2 dex in log $\dot{M}$), while CNO processing is 
relatively unimportant for the mass-loss rate.

The comparison of our predictions with the observed mass-loss 
behaviour of LBVs has shown that: 

\begin{enumerate}

\item The dominant driving mechanism of LBV winds is radiation pressure on spectral lines. 
       Other mechanisms are expected to play only a minor role.\\
\item The observed variable mass loss behaviour of LBVs of 
      up to $\sim$\,0.5 dex during their S\,Doradus-type variation cycles 
      can be explained in terms 
      of changes in the efficiency of line driving, as a result 
      of the recombination/ionisation of Fe\,{\sc iv}/{\sc iii} and 
      Fe\,{\sc iii}/{\sc ii}.\\
\item The \mdot(\teff) behaviour of AG\,Car can be matched to 
      within 0.1 dex when a shift in temperature of 
      $\Delta \teff = -6\,000$ K is applied to our 
      predictions. This shift is required to correct for
      an inaccurate calculation of the ionisation balance of 
      iron, as we have used a modified nebular approximation for this element. 
      Note that the shift is consistent with constraints
      set by observations of supergiants which indicate that the 
      Fe\,{\sc iv} to {\sc iii} and the Fe\,{\sc iii} to {\sc ii} 
      recombinations indeed occur at spectral types B1 and A0 
      (Lamers et al. 1995).\\
\item We have deduced a value of 35 \msun\ for the present-day mass 
      of AG\,Car. If we adopt a luminosity of 10$^{6}$ \lsun (which implies 
      an initial mass of $\sim$\,65\,\msun) and correct for 
      the amount of ejected material found in the circumstellar environment 
      of the star ($\sim$\,25\,\msun; Voors et al. 2000), we arrive at a present-day 
      mass upper limit of $\sim$\,40\,\msun, which is consistent with our finding of 
      35\,\msun.\\
\end{enumerate}



\begin{acknowledgements}

We thank Henny Lamers for many fruitful discussions.
JV acknowledges financial support from the Dutch NWO Council for Physical 
Sciences, the Particle Physics and Astronomy Research Council of 
the United Kingdom, and the Dutch research school for astrophysics (NOVA) for a travel grant.
AdK acknowledges support from NWO Pionier grant 600-78-333 to L.B.F.M.
Waters and from NWO Spinoza grant 08-0 to E.P.J. van den Heuvel.

\end{acknowledgements}

\end{document}